\documentclass[fleqn,10pt]{wlpeerj}

\usepackage{chemsym} 
\usepackage{hyperref}

\title{Linking thermodynamics and measurements of protein stability}

\author[1,*]{Kresten Lindorff-Larsen}
\author[1]{Kaare Teilum}
\affil[1]{Structural Biology and NMR Laboratory \& Linderstr{\o}m-Lang Centre for Protein Science, Department of Biology, University of Copenhagen, Copenhagen, Denmark}
\corrauthor[*]{Kresten Lindorff-Larsen}{lindorff@bio.ku.dk}

\begin{abstract}%
We review the background, theory and general equations for the analysis of
equilibrium protein unfolding experiments, focusing on denaturant
and heat-induced unfolding.
The primary focus is on the thermodynamics of reversible folding/unfolding
transitions and the experimental methods that are available for
extracting thermodynamic parameters. We highlight the importance of modelling
both how the folding equilibrium depends on a perturbing variable such as temperature
or denaturant concentration, and the importance of modelling the baselines in the
experimental observables.
\end{abstract}
\begin{document}
%\flushbottom
\maketitle
%\thispagestyle{empty}
%\newpage
% =======================================================
%           Introduction
% =======================================================
\section*{Introduction}
Protein folding is the spontaneous organisation of the polypeptide chain into a specific three dimensional structure. A detailed understanding of the interactions that determine protein structure and stability is key to our understanding of how proteins function and how we can engineer them for technological or medical purposes. A fundamental ingredient in studies of these properties is the ability to measure the stability of a protein, and indeed to define what we mean by stability. The ability to determine protein stability accurately is of central importance in protein engineering and design, as many proteins are engineered for improved thermostability. Further, measurements of (changes in) protein stability may be used to benchmark or even parameterize methods for computational design. In this review, we describe different approaches to probe protein stability. We focus in particular on the thermodynamic theories that underlie how protein stability varies with temperature and the concentration of chemical denaturants. We also focus on the historical developments that have led to current state of the art.

\section*{Conformations and states}
Due to the large dimension of configuration space (for example
defined by the position coordinates of all atoms) for a
protein, a very large number of possible conformations
exist. From an experimental point of view, it is impossible to
analyse all these conformations. Therefore, the first step in this
analysis is a reduction of the number of variables by grouping the
microscopic states (configurations) into larger states. The
following discussion is based on \cite{brandts1969conformational},
and takes as outset that we will probe some
physical or chemical parameter, $\alpha$, characteristic of the
system. $\alpha$ is an observable which we will use to
extract information about the system and could for example be the
result of a spectrometric measurement like absorption,
fluorescence or circular dichroism. We use
$\alpha_i$ to refer to the value for the parameter $\alpha$
for the microscopic state $i$. That is, $\alpha_i$ is the value of
$\alpha$ one would observe in the hypothetical situation where the
microscopic state $i$ is the only populated state. Finally, we
will assume that the macroscopic measurement of $\alpha$ termed
$\overline\alpha$, will be related to the microscopic $\alpha_i$
by a population-weighted average over all microscopic states:

\begin{equation}\label{eq:microaverage}
  \overline{\alpha} = \sum_{all\ microstates}p_i\alpha_i
\end{equation}
Here $p_i$ is the probability of being in the individual
microscopic state $i$. In general $\overline\alpha$ will be a
function of temperature ($T$), pressure ($P$) and the composition
of the system represented by $\xi$. Here, $\xi$ may for example include effects of added denaturants, but could also represent unfolding via pH \citep{tanford1961ionization} or alcohols \citep{miyawaki2011thermodynamic}.

The next step is to group microscopic states into larger states.
We will begin with a division of conformation space into two
subspaces $N$ and $U$ that do not overlap and together cover all relevant possible configurations. It
will be fruitful to think of $N$ and $U$ as consisting of native
and denatured microscopic states respectively, but so far the
equations are completely general leaving other possibilities open.
(We use $U$ for the denatured state to avoid confusion with the denaturant concentration
which we refer to below using $D$).
The macroscopic probability of being in $N$ ($U$) is given by the
sum of the probabilities for the microscopic states belonging to
$N$ ($U$):

\begin{equation}\label{eq:macroprobability}
  p_N = \sum_{\substack{states\\ in\ N}} p_i \quad , \quad%
  p_U = \sum_{\substack{states\\ in\ U}} p_j%
\end{equation}
The average of the property $\alpha$ in $N$ and $U$ is given by:

\begin{equation}\label{eq:averagealphas}
  \overline{\alpha}_N =%
    \frac{1}{p_N}\sum_{\substack{states\\ in\ N}}p_i\alpha_i \quad , \quad%
  \overline{\alpha}_U =%
    \frac{1}{p_U}\sum_{\substack{states\\ in\ U}}p_j\alpha_j
\end{equation}
Equations~\eqref{eq:microaverage} and
\eqref{eq:averagealphas} may be combined to express the observed
value of $\alpha$ in terms of the population and properties of the
two states $N$ and $U$, under the assumption that $\overline{\alpha}_N$ and $\overline{\alpha}_U$ are only weighted by the population to give the observed value:

\begin{equation}\label{eq:macroaverage}
  \overline{\alpha} = p_N \overline{\alpha}_N +%
    p_U \overline{\alpha}_U
\end{equation}
Remembering that the goal is to describe the thermodynamics of the
folding/unfolding reaction for a protein, we define an apparent
equilibrium constant, $K$, for the transition-reaction between the
$N$ and the $U$ region of conformation space by the equation:

\begin{equation}\label{eq:macroequil}
  K \equiv \frac{p_U}{p_N}
\end{equation}
Eq.~\eqref{eq:macroequil} describes how one can divide
conformation space into subspaces (here two) and defines an
equilibrium constant for the transition between them. For observables that
are population weighted, equations~\eqref{eq:macroaverage} and \eqref{eq:macroequil}
may be combined to provide an operational connection between $K$ and
the experimentally observable, $\overline\alpha$,
remembering that our definition of $N$ and $U$ ensures that $p_N + p_U = 1$:

\begin{equation}\label{eq:experimentalK}
   K(T,P,\xi) = \frac%
       {\overline{\alpha}_N(T,P,\xi) - \overline{\alpha}(T,P,\xi)}%
       {\overline{\alpha}(T,P,\xi) - \overline{\alpha}_U(T,P,\xi)}
\end{equation}
where the dependence of $T$, $P$ and $\xi$ has been written
explicitly. Eq.~\eqref{eq:experimentalK} is the theoretical
starting-point for most of the subsequent discussions. It states
that \emph{any} transition \emph{formally} can be considered a two-state
transition and shows how to determine the equilibrium constant $K$
for this transition experimentally. As discussed further below we stress, however, that
it is not advised to fit such a two-state model to a clear multi-state transition, since
the `baselines' $\overline{\alpha}_N$ or $\overline{\alpha}_U$ would then need to absorb
effects of transitions between states potentially leading to wrong thermodynamic and spectroscopic
parameters.

The above discussion provides a framework for much of the remainder of this review.
First, we note, however, that while $\overline{\alpha}(T,P,\xi)$ is an experimental
observable, the state-specific values $\overline{\alpha}_N(T,P,\xi)$ and $\overline{\alpha}_U(T,P,\xi)$
generally are not. These thus need to be modelled and determined through experimental procedures.
As we discuss in more detail below, this in turn generally requires one to build a model for how the
equilibrium constant $K$, or equivalently the unfolding free energy, $\Delta_rG_U$, varies as
one perturbs the experimental conditions such as the temperature, denaturant concentration, pressure or pH. In what follows, the state $N$ will generally refer to the native (folded) state and $U$ to an unfolded (denatured) state, but the discussion is general and the theory and equations may also be applied to other transitions (e.g. between a folded state and an intermediate) as long as they conform to the definitions above. We also note that the discussion and theory pertains to proteins that fold reversibly, and in general care should be taken to ensure that this is the case before analysing data in terms of thermodynamic models.

\section*{Measurement and extrapolation}
Looking more closely at Eq.~\eqref{eq:experimentalK} there are
three parameters on the right hand side defining the equilibrium
constant $K$ at some set of $T$, $P$ and $\xi$. The parameters are,
besides the directly observable $\overline{\alpha}$, the average
values of $\alpha$ in the two states $N$ and $U$ ($\overline{\alpha}_N$ and $\overline{\alpha}_U$). These are
generally not available and therefore Eq.~\eqref{eq:experimentalK}
does not directly make it possible to determine the stability, $K$, directly from the observable,  $\overline{\alpha}$.
In particular, it is clear that we need
some way of estimating $\overline{\alpha}_N$ and
$\overline{\alpha}_U$ in order to estimate $K$.

\begin{figure}[tbp]
  \centering
  \includegraphics[height=5cm]{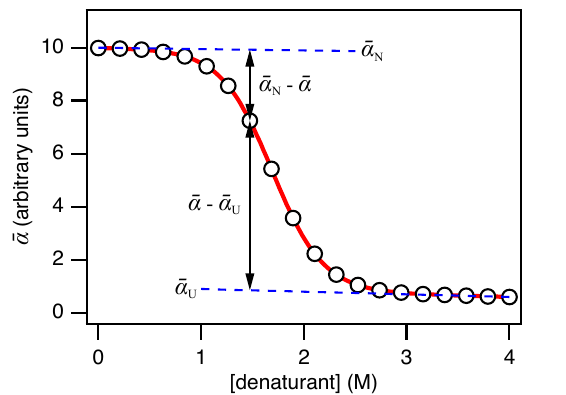}
  \caption{Extrapolation procedure for estimating $\overline{\alpha_N}$ and $\overline{\alpha}_U$. The white circles and red curve are (synthetic) experimental data. The dashed blue curves are estimates of $\overline{\alpha}_N$ and $\overline{\alpha}_U$ extrapolated from regions with almost pure $N$ and $U$ into the transition zone. The equilibrium constant is estimated as the ratio between the two distances $\overline{\alpha}_N - \overline{\alpha}$ and $\overline{\alpha} - \overline{\alpha}_U$. The curves were simulated using Eqs. \eqref{eq:N.baseline.D}, \eqref{eq:U.baseline.D} and \eqref{eq:alpha.vs.D} with $\overline{\alpha}_{N,0}$ = 10, $\beta_N$ = -0.05 M^{-1}, $\overline{\alpha}_{U,0}$ = 1, $\beta_N$ = -0.05 M^{-1}, $\Delta_rG_U^{\mathrm{H_2O}}$ = 16.9 kJ mol^{-1}, $m$ = 10 kJ mol^{-1} M^{-1} and $T$ = 298 K. A script used to generate the synthetic data is available at \url{https://github.com/KULL-Centre/papers/tree/master/2020/stability-lindorff-teilum}
  }
  \label{fig:extrapolate}
\end{figure}

In later sections of our review we describe in more detail how this may be done
specifically when inducing unfolding by temperature, denaturants or the two together, and refer the reader also to an earlier review for other practical considerations \citep{street2008protein}. Here, we first outline the general procedure for estimating the baselines:

\begin{enumerate}
  \item Vary one (or more) of $T$, $P$ and $\xi$ (the perturbing parameter(s)) until $p_N \approx 1$, thereby preparing almost pure $N$ (i.e. the folded state).
  \item Measure $\overline{\alpha}$ which will almost be equal 
        $\overline{\alpha}_N$ under these circumstances.
  \item Assume or derive some functional dependence of 
        $\overline{\alpha}_N$ on the perturbing parameter(s).
  \item Vary this (these) parameter(s) keeping $p_N \approx 1$.
  \item Extrapolate the behaviour of $\overline{\alpha}_N$ %
        to value of $T,\ P,\ \xi$ where it is needed using %
        the functional dependence and the experimental data.
\end{enumerate}
The procedure is repeated in a similar manner in order to estimate
$\overline{\alpha}_U$. See also Fig.~\ref{fig:extrapolate}.

In practice one often starts out with the protein as almost pure
$N$ (native protein, $p_N \approx 1$). As can be seen from
Eq.~\eqref{eq:macroaverage} this will in general mean that a
measurement of $\overline\alpha$ is a measurement of
$\overline\alpha_N$. The reason for this is that for most choices
of $\alpha$ the order of magnitude of $\overline\alpha_N$ and
$\overline\alpha_U$ are the same --- we are measuring rather small
differences\footnote{This is opposed to techniques like hydrogen
exchange where even though $p_N \approx 1$, one can still extract
information about $U$ since 
$\overline\alpha_N \ll \overline\alpha_U$}. Therefore one needs to
perturb the system by varying one of $T$, $P$ or $\xi$ in order to
get reasonable population of both states. This gives several new
problems the first one being that only $\overline{\alpha}$ is
experimentally accessible. Therefore one needs to use the
extrapolation procedure just outlined. The next problem is that
the conditions under which $K$ can be estimated are not usually
the same as those where the estimate is wanted. An example could
be that one has to raise the temperature to $50^\circ$C in order
to get significant amounts of $N$ and $U$ at the same time, but an
estimate of $K$ is needed at $25^\circ$C. The solution to this
problem is again extrapolation. In general one assumes or knows the
functional dependence of $K$ on the perturbing variable either
from theory or from previous experiments. This dependence is then
used to fit the data in the transition region and thereby provides
a way of extrapolating to the conditions where $K$ is needed.

The procedure outlined above was formalized by \cite{santoro1988unfolding} who suggested (focusing on denaturant-induced unfolding)
that Eq.~\eqref{eq:experimentalK} is rearranged to:

\begin{equation}\label{eq:santoro}
  \overline{\alpha}(T,P,\xi) = \frac%
                        {\overline{\alpha}_N(T,P,\xi) + \overline{\alpha}_U(T,P,\xi)K(T,P,\xi)}%
                        {1+K(T,P,\xi)}
\end{equation}
and that the experimental data consisting of measurements of
$\overline{\alpha}$ as a function of $(T,\ P,\ \xi)$ is fitted to
Eq.~\eqref{eq:santoro} by a non-linear least-squares procedure, as exemplified in Fig.~\ref{fig:extrapolate}. 
To implement it in practice, one needs to know or assume a specific functional dependence of
$\overline{\alpha}_N$, $\overline{\alpha}_U$ and $K$ on the
perturbing variable(s). As detailed with examples below, these dependencies may sometimes come from
basic thermodynamic theory, but also often contain some phenomenological component.
Using these
functional dependencies and the parameters estimated in the fitting
procedure one can evaluate $K$ at the desired value of the
perturbing parameter. Although this procedure is more rigorous than
using individual extrapolations for $\overline{\alpha}_N$,
$\overline{\alpha}_U$ and $K$ it must be stressed that the
procedure of evaluating $K$ using the fitted parameters may still
be an extrapolation \emph{in practice}. The procedure itself does
not change the fact that it is in general extremely difficult to get precise estimates
of $K$ in regions where $p_N$ is either 0 or 1. How the functional
dependencies of $\overline{\alpha}_N$, $\overline{\alpha}_U$ and
$K$ should be when the perturbing variable is $\xi$ (denaturant
induced unfolding) or $T$ (heat or cold induced unfolding) is
discussed in the sections below.

\noindent To summarize the whole procedure:
\begin{enumerate}
  \item Choose some physical or chemical observable $\alpha$ that %
        can be measured, and which differ in $N$ and $U$.
  \item Perturb the system by changing one or more of the variables %
        $T$, $P$ and $\xi$ in order to explore most of the region %
        of $p_U$ from 0 to 1.
  \item Measure $\alpha$ under all these conditions.
  \item Plot $\overline{\alpha}$ as a function of the perturbing %
        variable(s).
  \item Find functional dependencies of $\overline{\alpha}_N$, %
        $\overline{\alpha}_U$ and $K$ on the perturbing
        variable(s).
  \item Estimate the unknown parameters in these functions, generally by 
        non-linear least-squares regression of the data to Eq.~\eqref{eq:santoro}.
  \item Estimate $K$ at the value of $T$, $P$ and $\xi$ where it is needed.
\end{enumerate}

\section*{Denaturant induced unfolding}
Denaturants are compounds that are capable of destabilizing the
native structure of proteins relative to an unfolded or denatured ensemble of configurations. Two compounds in particular, urea
and guanidine hydrochloride (GdnHCl), are extensively used to study
protein stability and folding. Adding a denaturant in an
adequately high concentration to a protein solution will in most
cases cause unfolding of the protein. In the case of the
denaturants urea and GdnHCl, which typically cause unfolding at
rather high concentration (often several molar), a
special treatment is needed since the high concentrations used
require the compounds treated as co-solvents and not simply as
additives. In the rest of this text the term denaturant will
be used, but for all practical purposes we will have guanidinium salts or urea in mind.

The starting point for the discussion is
Eq.~\eqref{eq:santoro} and in particular the functional dependence
of $K$, $\overline{\alpha}_N$ and $\overline{\alpha}_U$ on $\xi$,
where $\xi$ in this section will be the amount of denaturant
present in the system parameterized either via the activity ($a_{denat.}$) or molar concentration ($D$). As discussed above it is necessary to know or assume functional dependencies of the perturbing variables in order to use Eq.~\eqref{eq:santoro} for analysing protein
stability measurements, and these dependencies will be
discussed below.

\subsection*{$K$ as a function of the amount of denaturant}
Several models for the effect of denaturants on protein stability
and how the dependence on $K$ should be treated have been
presented (reviewed in \cite{tanford1970protein,pace198614}).
As described below, most current research uses the so-called linear extrapolation method (LEM),
where the free energy of unfolding is assumed to depend linearly on the molar
concentration of the denaturant. We start, however, our discussion with models that are based on mechanistic ideas, and then show how they are related to the LEM.

The `denaturant binding model' \citep{brandts1964thermodynamicsII,aune1969thermodynamicsB}, uses the notion of linked functions \citep{wyman1964linked}.
The theory describes how the equilibrium constant for a reaction,
\mbox{$N \rightleftarrows U$}, will be influenced by the presence
of a ligand, $x$, which will bind to $N$ and/or $U$. The reaction
studied is the denaturation process and the ligand is the
denaturant. The result is that the equilibrium constant for the
unfolding reaction, $K_U$, is related to the number of denaturant
molecules taken up during the reaction, $\Delta \nu$, by:

\begin{equation}\label{eq:linkage}
    \left( \frac{\partial \ln K_U}{\partial \ln a_{denat.}}\right)_{T,\ pH} = %
        \Delta \nu
\end{equation}
where $a$ denotes activity. As the name implies, the model
implicitly assumes a binding of denaturant molecules to the
protein. Assuming there are $n_U$ `binding sites' on the denatured
(unfolded) protein and $n_N$ on the native protein and that each
of these binding sites are independent and characterized by an individual
intrinsic (microscopic) binding constant, $k_i$, standard theory
for multiple binding sites \citep{tanford1961ionization} leads to:

\begin{equation}\label{eq:multiplebinding}
  K_U(a_{denat.}) = K_{U,0} \frac%
    {\prod_{i=1}^{n_U}(1+k_{i,U}a_{denat.})}%
    {\prod_{i=1}^{n_N}(1+k_{i,N}a_{denat.})}
\end{equation}
Here $K_{U,0}$ denotes the value of $K_U$ in the absence of
denaturant. Eq.~\eqref{eq:multiplebinding} contains too many
unknowns to be estimated from a typical experiment. Therefore
further approximations are needed. If all sites on $N$ are
considered identical, then all $k_{i,N} = k_N$. If
also all $k_{i,U}$ are set equal to $k_U$ this leads to:

\begin{equation}\label{eq:bindingtanford}
  K_U(a_{denat.}) = K_{U,0} \frac%
    {\left( 1+k_Ua_{denat.}\right) ^{n_U}}%
    {\left( 1+k_Na_{denat.}\right) ^{n_N}}
\end{equation}
Even this equation with five unknown parameters may be too
complicated to analyse using normal denaturation data. The model may be simplified further by
assuming that the difference in binding of denaturant to
$N$ and $U$ lies in the number of sites and not in the individual
affinities to the binding sites. This leads to:

\begin{equation}\label{eq:bindingsimple}
   K_U(a_{denat.}) = K_{U,0}%
     \left( 1+ka_{denat.}\right) ^{\Delta n}
\end{equation} 
or the equivalent:

\begin{equation}\label{eq:bindingsimple.deltaG}
  \Delta_rG_U(a) = \Delta_rG_U^{\mathrm{H_2O}}-\Delta n RT\ln(1+ka_{denat.})
\end{equation}
where $\Delta n = n_U - n_N$ and $\Delta_rG_U^{\mathrm{H_2O}}=-RT\ln K_{U,0}$.
In one study, \cite{aune1969thermodynamicsB} studied GdnHCl denaturation of lysozyme and found that the model described the data using $\Delta n =
7.8$ and $k = 3.0$ indicative of weak binding. Values for $k$ and
$\Delta n$ are discussed further in \cite{pace198614}.
On the other hand, \cite{brandts1964thermodynamicsI,brandts1964thermodynamicsII} found that this equation could not
explain his data on urea denaturation of chymotrypsinogen. 

By replotting the data measured by \cite{aune1969thermodynamicsA,aune1969thermodynamicsB}, \citet[Fig. 8]{tanford1970protein}  showed a plot of $\log K_U$
\emph{vs.} GdnHCl concentration. $K_U$ was measured in an interval
of 2--4~M GdnHCl and showed a linear dependence of $\log K_U$ on
[GdnHCl]. This prompted \cite{greene1974urea} to suggest what 
later became known as the LEM \citep{pace2000linear} in
which $K_U$ is related to the amount of denaturant present through
the equation:

\begin{equation}\label{eq:LEM}
  \Delta_rG_U(D) = \Delta_rG_U^{\mathrm{H_2O}}-mD
\end{equation}
In this equation, $D$ is the molar concentration of denaturant and
$m$, the so called $m$-value, is the (negative) slope in a $
\Delta_rG_U$ \emph{vs.} $D$ plot (the convention of defining $m$ as the negative slope ensures that $m$ is a positive quantity). This method, which today
has become the standard method \citep{maxwell2005}, is mainly used because of its
success in describing experimental data. It is simple and describes
$\Delta_rG_U(D)$ by only two parameters as compared to three in
Eq.~\eqref{eq:bindingsimple.deltaG} or five in
Eq.~\eqref{eq:bindingtanford}. The major drawback of the method is
that it is not accompanied by a theory, leaving interpretations of
the two parameters $\Delta_rG_U^{\mathrm{H_2O}}$ and the $m$-value more
difficult. To the extent that it is an accurate phenomenological description
of protein denaturation, the fitted values of $\Delta_rG_U^{\mathrm{H_2O}}$ represent accurately
the stability of a protein extrapolated to the absence of denaturant. From a mechanistic
perspective, it is unclear what the $m$-value represents e.g. in terms of the preferential
interaction with unfolded proteins, but further phenomenological observations
suggest that it is correlated with the change in solvent accessible surface area during
unfolding \citep{schellman1978solvent,myers1995denaturant,geierhaas2007bppred}.

As discussed above,
$\Delta_rG_U$ can only be determined in a limited interval around
zero where both the folded and denatured states have sizable populations. This means that the value of $\Delta_rG_U^{\mathrm{H_2O}}$ in
Eq.~\eqref{eq:LEM} is an \emph{extrapolated} value. For most
proteins denaturation is observed with $D>1~M$ which means that
although $\Delta_rG_U$ may be a linear function in $D$ when it can
be measured, it may still behave differently at lower denaturant
concentrations. Therefore Eq.~\eqref{eq:LEM} may better be interpreted
as a truncated series-expansion of $\Delta_rG_U$ in $D$ around the
value $D_{50}$, which is defined as the value of $D$ which gives
$\Delta_rG_U=0$:

\begin{equation}
  \begin{split}\label{eq:deltaGseries}
  \Delta_rG_U(D) = & \left. \frac{\mathrm{d} \Delta_rG_U}%
    {\mathrm{d} D}\right| _{D_{50}}\mspace{-18.0mu} (D-D_{50}) \ +\ %
  \left. \frac{\mathrm{d} ^2 \Delta_rG_U}{\mathrm{d} D^2}\right| _{D_{50}}%
    \mspace{-18.0mu}\frac{(D-D_{50})^2}{2!}\\ %
  &+ \cdots +\ %
    \left. \frac{\mathrm{d} ^{n-1} \Delta_rG_U}{\mathrm{d} D^{n-1}}\right| _{D_{50}}%
    \mspace{-18.0mu}\frac{(D-D_{50})^{n-1}}{(n-1)!}\ +\  R_n
  \end{split}
\end{equation}
Assuming the expansion is represented well by the first-order term
alone in some sufficient $D$-interval around $D_{50}$ one finds:

\begin{align}
  m &\approx - \left. \frac{\mathrm{d} \Delta_rG_U}{\mathrm{d} D}\right| _{D_{50}}%
    \label{eq:mvalue.definition}\\%
  \Delta_rG_U(D) &\approx -m(D-D_{50})%
    \label{eq:alternativeLEM}
\end{align}
Using these equations, the extrapolated value
$\Delta_rG_U^{\mathrm{H_2O}}$ in Eq.~\eqref{eq:LEM} is related to $m$ and
$D_{50}$ by:

\begin{equation}\label{eq:LEMdeltaG}
  \Delta_rG_U^{\mathrm{H_2O}} \approx mD_{50}
\end{equation}
assuming that the expansion in Eq.~\eqref{eq:deltaGseries} can be
approximated well by the truncated form
Eq.~\eqref{eq:alternativeLEM}. Other authors \emph{define} the $m$-value
by:

\begin{equation}\label{eq:alt.mvalue.definition}
  m \equiv - \frac{\mathrm{d} \Delta_rG_U}{\mathrm{d} D}
\end{equation}
which in principle leaves the $m$-value as a function of $D$. In the case where
Eq.~\eqref{eq:LEM} is valid in the whole $D$-interval, these two
definitions (equations \eqref{eq:mvalue.definition} and
\eqref{eq:alt.mvalue.definition}) will obviously coincide. There is, however, some evidence to suggest that the slope of $\Delta_rG_U(D)$-vs-$D$ may not be constant, i.e. the LEM does not hold fully, \citep{yi1997characterization,amsdr2019urea}, which may explain why extrapolated values of $\Delta_rG_U^{\mathrm{H_2O}}$ may differ between different denaturants \citep{moosa2018denaturant}.

To accommodate different models and to put equations like
\eqref{eq:LEM} on a more solid theoretical ground, a general thermodynamic theory for denaturant
induced unfolding is needed. This work was begun by \cite{schellman1978solvent,schellman1987selective,schellman1990fluctuation,schellman1994thermodynamics,schellman1996enthalpy,schellman2002fifty,schellman2003protein}
who developed a theory for what he called
`solvent denaturation'. The reader is referred to these 
papers for a full description, and only the main results and
interpretations will be summarized here. The starting point is the
denaturant binding model and the theory of multiple binding sites.
From a theoretical point of view one criticism of that model
is, that it assumes there are binding sites and treats those
binding sites in the language of standard thermodynamic binding
theory. This may be correct for ligands with high affinity, but for
`ligands' like urea or GdnHCl which exert their effect at very
high concentration and therefore must be classified as weak
binders, the binding theory misinterprets the notion of binding.
Or in the words of \cite{schellman1987selective}:
\begin{quote}
{\em The definition of binding then comes into question. If a site
is occupied by a denaturant molecule, is it `bound' or not?
Thermodynamically, the answer is `yes' only if the binding is in
excess of expectation on the basis of solvent composition.}
\end{quote}
Therefore Schellman starts out to redefine the notion of binding
in a thermodynamic formulation. This is done using multicomponent
solution thermodynamics \citep{casassa1964thermodynamic}, statistical
mechanic fluctuation theory \citep{fowler1936statistical} and refining the
linkage theory of \cite{wyman1964linked} to include weakly binding ligands.
The observation is that using this new general notion of
binding, almost all the equations of standard binding theory can
be recovered in the same mathematical form but with new
interpretations.

Using the Scatchard notation \citep{scatchard1946physical} in a molal basis,
the chemical potential of component $j$ in the solution is given
by:

\begin{equation}\label{eq:ScatchardChemPot}
  \mu_j = \mu_{0j}^\ominus + RT\ln (m_j/m^\ominus) + RT\beta_j%
  \qquad j=2,3
\end{equation}
Here $\mu_{0j}^\ominus$ is the chemical potential of the reference
state, $m_j$ is the molality and $RT\beta_j$ is the excess free
energy. We here use the notation with $j=1$ for the principal solvent
(water), $j=2$: for the protein and $j=3$: for the co-solvent
(denaturant). The equations can easily be
generalized to more than one co-solvent. With this notation, Eq.~\eqref{eq:linkage} then takes the form:

\begin{align}
  \Gamma_{32} &\equiv \lim_{m_2 \to 0} \left(\frac%
    {\partial m_3}{\partial m_2}\right)_{\mu_3} \label{eq:new.binding}\\%
  \left(\frac{\partial \ln K_U}{\partial \ln a_3}\right)%
    &= \Delta \Gamma_{32} \label{eq:new.linkage}
\end{align}
where Eq.~\eqref{eq:new.binding} is the thermodynamic definition
of binding and the $\Delta$ in Eq.~\eqref{eq:new.linkage} refers
to the difference in this `binding' between the native and the
denatured states. \cite{casassa1964thermodynamic} have shown, that
$\Gamma_{23}$, apart from a negligible term, is the binding
quantity measured in equilibrium dialysis; if one mole of protein
is added at one side of the dialysis membrane then $\approx
\Gamma_{23}$ moles of co-solvent have to be added in order to keep
its chemical potential constant. Eq.~\eqref{eq:new.linkage}
shows how the equilibrium constant for a conformational change
depends of the thermodynamics of binding/interactions in the two
conformations. Although the \emph{mathematical} form of equations
\eqref{eq:linkage} and \eqref{eq:new.linkage} are the same, they
have different interpretations due to the new definition of
binding. To look at an example, assume a single site reaction
like:

\begin{center}
hydrated~site~+~denaturant~=~site$\cdot$denaturant~+~water\\%
\end{center}
For this reaction standard binding theory would yield the binding
isotherm in Eq.~\eqref{eq:standard.gamma} whereas the solvent
exchange model yields Eq.~\eqref{eq:new.gamma}:

\begin{align}
  \Gamma_{32} &= \frac{K\mathrm{[denaturant]}}{1+K\mathrm{[denaturant]}}%
    \label{eq:standard.gamma}\\%
  \Gamma_{32} &= \frac{(K'-1)\chi_3}{1+(K'-1)\chi_3}%
  \label{eq:new.gamma}
\end{align}
where $K$ is the (molar) equilibrium constant, $\chi_3$ is the
mole fraction of co-solvent (denaturant) and $K'$ is equal to
$Kf_1/f_3$ where $f_j$ are activity coefficients on the mole
fraction scale. In the standard binding model $K$ will always be
positive whereas the \emph{effective} binding constant in
Eq.~\eqref{eq:new.gamma}, $K'-1$, will be negative for situations
with preferential hydration. This is the main result of the model.
It can distinguish the concept `binding' from the concept
`occupation' and gives a way of analysing this new definition of
binding. Occupation within an area (a site) on a protein, only
constitutes binding if the co-solvent is occupying that area more
than would be predicted from solvent composition alone.

Schellman's solvent denaturation model leads to a well-defined theory that links denaturant concentrations (or activities and mole fractions) to preferential binding and thus how a folding equilibrium might be affected. Nevertheless, the resulting equations do not directly lead to something like the LEM, which is generally assumed to describe experiments rather well. Through a number of assumptions
\cite{schellman1987selective} arrives at an effective LEM where $\Delta_rG_U$
is a linear function of denaturant mole fraction, but this model is not commonly used to describe stability measurements.
Defining $\beta_{23}=(\partial \beta_2/\partial m_3)$ (see
also Eq.~\eqref{eq:ScatchardChemPot}) and letting the $\Delta$
operator define a difference between the unfolded and native
state, the task of showing the validity of a LEM is equivalent to
showing that $\Delta \beta_{23}$ is linear in denaturant
concentration.

In theory there are no reasons why a LEM theory should use the
\emph{molarity} of denaturant instead of other concentration
scales. To address this question \cite{schellman1990fluctuation} tested linear
extrapolations of $\Delta_rG_U$ in activity, molarity and molality
for four proteins. The extrapolated value for
$\Delta_rG_U^{\mathrm{H_2O}}$ was compared to that estimated from
calorimetric experiments. The result was that the estimate using
molarity and molality had the best agreement with the calorimetric
estimate. Linearity of the plots were tested with a $\chi^2$ test
and showed that overall, the molarity scale gave plots which were
most linear, providing a practical argument why the LEM is formulated as it is.

To summarize this discussion, we note that standard protein
denaturation experiments contain relatively little information about the thermodynamics of unfolding.
This in turn implies that only a
few parameters can be extracted, and so simple models and equations are generally used. Of the equations presented, `the linear extrapolation
method', as presented in equations \eqref{eq:LEM} and
\eqref{eq:alternativeLEM}, has been the most successful. Indeed, even with only two parameters, the LEM can show substantial parameter correlation during non-linear regression \citep{lindorff2019dissecting}.  At the
moment there is no good theoretical explanation why these simple
equations seem to describe most of the protein denaturation data
so well. The solvent exchange model is a big step towards such an
explanation, but presumably new types of experiments have to be
designed in order to assign values to the parameters in the model.
Analysis of enzyme kinetics in the presence of urea \citep{wu1999new} have provided additional information regarding
interaction between proteins and denaturants. Also, extensions of the solvent exchange model have been proposed  \citep{amsdr2019urea}, but are not discussed further here.

\subsection*{$\overline\alpha_N$ and $\overline\alpha_U$ as functions of denaturant concentration}
We now return to the overall strategy as described in the introduction and captured in Eq.~\eqref{eq:santoro}.
We remind ourselves that we not only need a parametric model for how the free energy of unfolding (or $K(T,P,\xi)$)
depends on the solvent composition ($\xi$) (such as the LEM described above), but also need to model
how the value of the experimental observable ($\overline{\alpha}$) varies in the two states.

How the average value of the observable
$\alpha$ in states $N$ and $U$ depends on the the solvent composition $\xi$ (e.g. the denaturant concentration),
will of course depend on which
observable has been chosen. In general there are no good theories
(like thermodynamics) to use when one looks for the required
functional dependencies. For this reason one will in general be
satisfied by empirically determined functions. For fluorescence
and absorption spectroscopy \cite{schmid1997optical} provides plots of
the absorption and fluorescence free tyrosine and tryptophan as a function
of denaturant concentration (urea: 0--8.5~M, GdnHCl: 0--7~M). In
the cases where an effect is observed, there is an increase in
absorption/fluorescence. The results are summarized in
Table~\ref{tab:abs.fluor}.

\begin{table}[tbp] 
   \begin{center}
   \caption{Effect of denaturants on tyrosine and tryptophan %
            absorption and fluorescence. Values presented are the %
            percent increase in absorption/fluorescence when %
            going from water to a solution containing high denaturant %
            concentration (8.5~M urea or 7.0~M GdnHCl). Data taken
            from  \cite{schmid1997optical}, where experimental conditions can
            be found.}
   \label{tab:abs.fluor}
   \begin{tabular}{l c p{0.0cm} c p{0.1cm} c p{0.0cm} c}
   \hline %
      &  \multicolumn{3}{c}{Tyrosine} & & \multicolumn{3}{c}{Tryptophan} \rule[-4pt]{0pt}{18pt} \\ %
                \cline{2-4}                              \cline{6-8}\\
      \vspace{-20pt}\\
           &    Absorption    &   &   Fluorescence \rule[-4pt]{0pt}{16pt}   &   &   Absorption    &   &   Fluorescence\\ %
                \cline{2-2}     \cline{4-4}              \cline{6-6}     \cline{8-8}\\ %
      \vspace{-20pt}\\
      Urea &    50\%    &   &   20\%             &   &   20\%    &   &   40\%  \rule{0pt}{12pt} \\ %
      GdnHCl&   40\%    &   &   10\%             &   &   60\%    &   &   10\%  \rule[-6pt]{0pt}{18pt} \\ %
      \hline
   \end{tabular}
   \end{center}
\end{table}

The indicated percentages are the relative increases in
absorption/fluorescence intensity. If free tyrosine and tryptophan
are good models for protein fluorescence and absorption then these
results indicate that a linear function should be adequate for a
description of $\overline\alpha_N$ and $\overline\alpha_U$ as a
function of denaturant concentration. In practice one observes
both increases and decreases of fluorescence and absorption in the
pre- and post-transition baselines in protein denaturation
experiments. This is not surprising since for example the
fluorescence intensity of a protein is not simply a sum of the
fluorescence from its amino acid constituents in their free states. Other effects like
quenching by nearby amino acids or fluorescence resonance energy
transfer between tryptophan and tyrosine complicates predictions of how $\overline\alpha_N$ and
$\overline\alpha_U$ should be modelled \citep{vivian2001mechanisms,callis2011predicting}. Instead, a simple trial and error approach is therefore generally
followed.  A good starting point will simply be a constant. If the
fit is not satisfactory, then using a straight line will in most
cases be sufficient. If this is not good enough one should then
increase the complexity of the function until an adequately (by
some statistical criterion) good fit is obtained. One key problem is that
it is only easy to observe deficiencies of the baselines near the ends where the pure folded or unfolded states
are populated, but on the other hand, the accuracy of the fit to the thermodynamic model requires that
the baselines are also accurate near the transition region.

Linear baselines are used in most cases where denaturant induced unfolding has been measured by CD or fluorescence spectroscopy:

\begin{equation}\label{eq:N.baseline.D}
    \overline{\alpha}_N\left(D\right)=\overline{\alpha}_{N,0}+{\beta}_{N}D
\end{equation}
\begin{equation}\label{eq:U.baseline.D}
    \overline{\alpha}_U\left(D\right)=\overline{\alpha}_{U,0}+{\beta}_{U}D
\end{equation}
Inserting these baselines into Eq.~\eqref{eq:santoro} and using $K=\exp(-(\Delta rG_U^{\mathrm{H_2O}}-mD)/RT)$ from Eq.~\eqref{eq:LEM} we arrive at:
\begin{equation}\label{eq:alpha.vs.D}
    \overline{\alpha}\left(D\right)=\frac{\left(\overline{\alpha}_{N,0}+{\beta}_{N}D\right)+\left(\overline{\alpha}_{U,0}+{\beta}_{U}D\right)\exp{\left(\frac{\Delta_rG_U^{\mathrm{H_2O}}-mD}{RT}\right)}}{1+\exp{\left(\frac{\Delta_rG_U^{\mathrm{H_2O}}-mD}{RT}\right)}}
\end{equation}
which is widely accepted as the standard for analysis of denaturant induced protein unfolding \citep{maxwell2005}. Simulated data using this equation are show in Fig.~\ref{fig:extrapolate}.

\section*{Heat induced unfolding}
In contrast to denaturant induced unfolding there already exists a
well tested theoretical apparatus for describing the
thermodynamics of heat and cold denaturation of proteins. This
model, which is based on standard thermodynamics, has already
been discussed extensively in the literature \citep{privalov1979stability,becktel1987protein,privalov1988stability} and will not
be repeated here. Only the main features and the standard
approximations will be discussed. When heat stability is measured
by calorimetry the thermodynamic equations are the only equations
needed. On the other hand, if a spectrometric technique like
fluorescence is used, one has to use Eq.~\eqref{eq:santoro} in
order to analyse the data. Here thermodynamics will show you how
to setup a function for $K(T)$ but it will not tell you how
$\overline\alpha_N(T)$ and $\overline\alpha_U(T)$ should be
modelled. The question of finding appropriate functions for
$K(T)$, $\overline\alpha_N(T)$ and $\overline\alpha_U(T)$ are covered in the following two sections.

\subsection*{$K$ as a function of temperature}
Below we continue to analyse the thermodynamics of unfolding under the assumption
that a two-state description is sufficient. As above, we examine only reversible transitions, and note that in particular thermal unfolding may often be irreversible. For a reversible two-state transition, the relevant equations for discussing the
form of $K(T)$ are:

\begin{align}
  \Delta_rG_U(T)    &= -RT\ln K_U(T)%
    \label{eq:deltaG.of.K}\\%
  \Delta_rG_U(T)    &= \Delta_rH_U(T) - T\Delta_rS_U(T)%
    \label{eq:deltaG.of.H.and.S}\\%
  \left(\partial \Delta_rH_U(T)/\partial T\right)_P%
    &= \Delta_rC_{p,U}(T)
    \label{eq:heat.capacity.def}\\%
  \left(\partial \Delta_rS_U(T)/\partial T\right)_P%
    &= \Delta_rC_{p,U}(T)/T
    \label{eq:entropy.of.T}\\%
  \left(\partial \Delta_rG_U(T)/\partial T\right)_P%
    &= - \Delta_rS_{U}(T)
    \label{eq:deltaG.of.T}\\%
  \left(\mathrm{d} \ln K_U^\ominus(T) / \mathrm{d} (T^{-1})\right) &=%
    - \left(\mathrm{d} \left(\Delta_rG_U^\ominus(T)/RT\right) / \mathrm{d} (T^{-1})\right)%
    \notag \\%
  &= - \Delta_rH_U^\ominus(T)/R
    \label{eq:vant.hoff}%
\end{align}

\begin{figure}[tbp]
  \includegraphics[height=5cm]{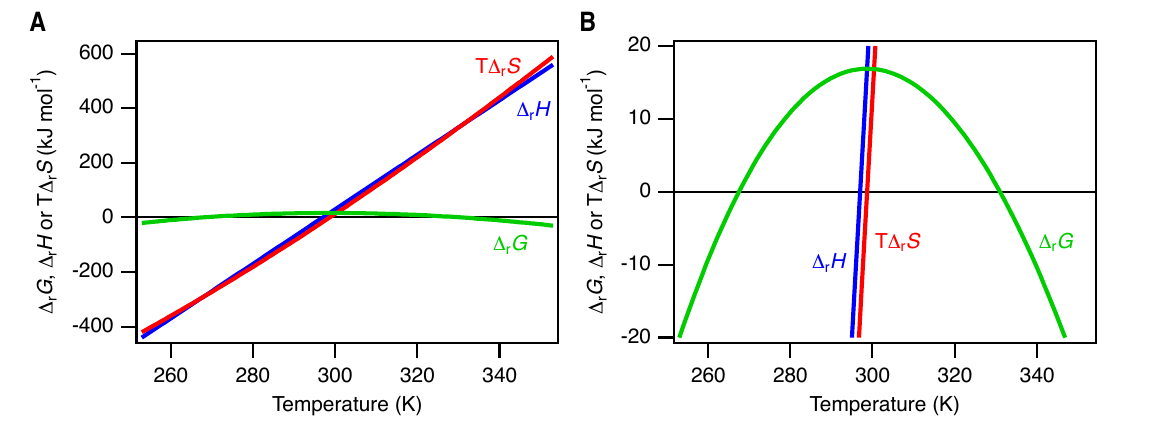}
  \caption{
            Synthetic data for the temperature dependency of thermodynamic parameters. Blue line: $\Delta_rH$. Red line: $T\Delta_rS$. Green line: $\Delta_rG$. A) shows the full energy interval for $\Delta_rH$ and $T\Delta_rS$. B) is a close-up on $\Delta_rG$. The data were modelled assuming a temperature independent $\Delta_rC_p$ = 10 kJ mol^{-1}K^{-1}, $T_m$ = 331 K (58 $^\circ$C) and $\Delta_rH$ = 340 kJ mol^{-1} at the melting temperature.}
  \label{fig:temperature-profile}
\end{figure}

One of the defining features when looking at protein
denaturation thermodynamics is the large and positive change in
heat capacity that is generally found to accompany the unfolding  process. This
is such a general phenomenon that it has been suggested as the
definition of protein denaturation \citep{lumry1969thermodynamic}. The
exact reasons for this large $\Delta_rC_{p,U}$ are uncertain but
they are related to the so called `hydrophobic effect', a term
with so many definitions that a full discussion is beyond the
scope of the present discussion. We shall therefore simply note
that the effect is related to the specific properties of water as a
solvent for hydrophobic compounds, and further point the reader to
discussions elsewhere \citep{privalov1989hydrophobic,muller1990search,lins1995hydrophobic,ben2016water}.
The effect of having a large
$\Delta_rC_{p,U}$ can be seen from equations
\eqref{eq:heat.capacity.def} and \eqref{eq:entropy.of.T}, which show that both the enthalpy and the entropy change will
be highly temperature dependent.
At most temperatures $|\Delta_rG_U|$ is much smaller than either
of $|\Delta_rH_U|$ and $|T\Delta_rS_U|$ (Fig.~\ref{fig:temperature-profile}A). This means that the
stability of the protein is a small number compared to the two, generally
opposing, effects of enthalpy and entropy change.

It is also generally the case that $\Delta_rC_{p,U} > |\Delta_rS_U|$. To
see what this means, one can use Eq.~\eqref{eq:deltaG.of.H.and.S}
instead Eq.~\eqref{eq:deltaG.of.T} to calculate the temperature
dependence of $\Delta_rG_U$:

\begin{equation}\label{eq:alt.deltaG.of.T}
  \begin{split}
  \left(\partial \Delta_rG_U(T)/\partial T\right)_P%
   &= \left(\partial \Delta_rH_U(T)/\partial T\right)_P - %
   \left(\partial (T\Delta_rS_U(T))/\partial T\right)_P\\
   &= (\Delta_rC_{p,U}(T)) - (\Delta_rC_{p,U}(T)+\Delta_rS_U(T))%
   = -\Delta_rS_U(T)
   \end{split}
\end{equation}
Again, the results show that $\Delta_rG_U$ is much less temperature
dependent than either of $\Delta_rH_U$ and $T\Delta_rS_U$ --- the
increase in denaturation enthalpy by increasing T is almost
exactly compensated by an increase in $T\Delta_rS_U(T)$, leaving
only a small temperature effect (of either sign) on
$\Delta_rG_U(T)$.

We proceed by examining the curvature of how $\Delta_rG_U$ changes as a function of $T$:

\begin{equation}\label{eq:deltaG.curvature}
  \left( \partial ^2\Delta_rG_U(T)/\partial T^2\right)_P%
    = - \Delta_rC_{p,U}/T 
\end{equation}
which in general is negative at all relevant temperatures. Generally, there exists some temperature, which we term $T_S$, where
$\Delta_rS_U=0$, and so $\Delta_rG_U$ will be at a maximum at $T_S$.
Going both up and down from this temperature will decrease the stability ($\Delta_rG_U$). To sum up,
$\Delta_rG_U$ is a small number compared to $\Delta_rH_U$ and
$T\Delta_rS_U$ and is much less temperature dependent than either.
$\Delta_rG_U(T)$ has a maximum at $T_S$, and below and above this temperature the stability will decrease (Fig.~\ref{fig:temperature-profile}B).

The behaviour described above are those one must bear in mind when choosing
some appropriate function $K(T)$. Obviously one cannot neglect the
temperature dependence of $\Delta_rH_H$ and $\Delta_rS_U$. The
simplest function that will include these effects is one using a
temperature independent heat capacity change. This approximation
will in many cases be adequate for describing experimental data,
and only with very precise measurements may it be possible 
to observe deficiencies arising from it. Assuming
$\Delta_rC_{p,U}(T)=\Delta_rC_{p,U}$ one can integrate equations
\eqref{eq:heat.capacity.def} and \eqref{eq:entropy.of.T} and
together with Eq.~\eqref{eq:deltaG.of.H.and.S} one arrives at:

\begin{equation} \label{eq:constant.deltaCp}
  \begin{split}
  \Delta_rG_U(T) = &\Delta_rH_U(T_0) - T\Delta_rS_U(T_0) +\\%
    &\Delta_rC_{p,U}(T - T_0 - T \ln(T/T_0))%
  \end{split}
\end{equation}
where $T_0$ is some reference temperature. Choosing this
temperature as the melting temperature, that is the (highest)
temperature where $\Delta_rG_U(T)=0$ the equation can be reduced
to:

\begin{align}
  \Delta_rG_U(T) = &\Delta_rH_U(T_m)(1-T/T_m) + \notag\\%
    & \Delta_rC_{p,U}(T - T_m - T \ln(T/T_m))\label{eq:constant.deltaCp.2}\\%
\intertext{using that:}
  \Delta_rG_U(T_m) = &\Delta_rH_U(T_m) -T_m \Delta_rS_U(T_m) = 0%
  \label{eq:deltaG.at.Tm}
\end{align}
Eq.~\eqref{eq:constant.deltaCp.2} contains three unknown
parameters: $\Delta_rH_U(T_m)$, $\Delta_rC_{p,U}$ and $T_m$. If
the experimental data does not contain enough information to allow
the estimation of these three parameters, then the further (generally poorer)
approximation that $\Delta_rC_{p,U}=0$ will lead to:

\begin{equation}\label{eq:zero.deltaCp}
  \Delta_rG_U(T) = \Delta_rH_U(T_m)(1-T/T_m)
\end{equation}

We note also that if the goal is to determine changes in protein stability at room temperature, then a theoretically-motivated expression can be used to estimate this from changes in $T_m$ \citep{watson2017size}.

\subsection*{$\overline\alpha_N$ and $\overline\alpha_U$ as functions of temperature}
%\label{sec:alpha(T)}%
With a set of equations in hand that describe how $K$ depends on temperature, we proceed to examine the
baselines in heat-induced unfolding experiments.
The temperature effect on spectrometric techniques can in general
be difficult to predict and will depend on which technique is
used. The primary focus in this section will be the
temperature dependence of fluorescence intensities, and will exemplify how
one may model the baselines using either physical principles or fully empirical approaches.

For both fluorescence \citep{bushueva1978relationship,eftink1994use} and absorption
\citep{sinha2000possible} spectroscopy it has been observed that a plot
of $\overline\alpha$ \emph{vs}. $T$ can be non-linear even outside
the transition region. This effect must be incorporated into the
expressions for $\overline\alpha_N(T)$ and $\overline\alpha_U(T)$.
At the same time one wants to use as few parameters as possible in
Eq.~\eqref{eq:santoro}. Depending on the quality and amount of
baseline data one should (in general) not use more than three and
preferably less free parameters for determining each baseline. If
one is to incorporate all these preferences into a model of the
baseline, this is going to be more difficult than in the case of
denaturant effects, which are often described well by constant or linear baselines.
In particular one needs a simple model that
incorporates non-linear baselines. Below we describe one such a model for
the temperature-dependency of protein fluorescence based on \cite{bushueva1978relationship}.

In general the fluorescence intensity, $F$, is given by the
expression:
\begin{equation}\label{eq:fluorescence}
  F = z I_{abs.} q
\end{equation}
where $z$ is a factor that among other things is related to the
apparatus, $I_{abs.}$ is the amount of absorbed light and $q$
is the quantum yield --- the number of photons emitted per photon
absorbed. The starting point for the model we describe is that all temperature
dependence of $F$ is modelled to be contained in the
temperature dependence of the quantum yield. For a single
chromophore, $q$ is related to the rate constants for the
emission, $k_F$, and the non-emitting processes (first-order or
pseudo-first-order), $k_i$, by the expression:

\begin{equation}\label{eq:quantum.yield}
  q = \frac{k_F}{k_F+\sum k_i}
\end{equation}
Normalizing all rate constants by $k_F$ and grouping the
non-radiative processes into temperature dependent $k_{j,T}$ and
non-temperature dependent $k_{i,NT}$, Eq.~\eqref{eq:quantum.yield}
can be rewritten:
\begin{equation}\label{eq:quantum.yield.temp}
  q(T) = \frac{1}{1 + \sum_i k_{i,NT} + \sum_j k_{j,T}}%
\end{equation}
The task is then to find some expression for the temperature
dependence of $\sum_j k_{j,T}$. Bushueva \emph{et al}. suggests
that this can be done by using some general temperature dependence
function, $f(T)$, for all $k_{j,T}$:

\begin{equation}\label{eq:quantum.yield.general.temp}
  q(T) = \frac{1}{1 + \sum_i k_{i,NT} + f(T) \sum_j k_{j,\,eff}}%
\end{equation}
Here $\sum_j k_{j,T}$ has been split into a the temperature
dependent $f(T)$ and the temperature independent $\sum_j
k_{j,\,eff}$. Finally it is suggested that the form of $f(T)$ is
given by the temperature dependence of a diffusion-limited
reaction. For such a reaction every formation of an encounter pair
leads to a reaction. From Fick's first law of diffusion and the
Einstein-Stokes relationship one can determine that the rate constant
for a diffusion-limited reaction is given by \citep{steinfeld1989chemical}:

\begin{equation}\label{eq.diff.limited}
  k_D \approx \frac{2k_B(r_A+r_B)^2T}{3r_Ar_B\eta(T)}
\end{equation}
where $r_A$ and $r_B$ are the radii of the two reacting species
and $\eta(T)$ is the (temperature dependent) viscosity. Grouping
all constants and combining equations \eqref{eq:fluorescence},
\eqref{eq:quantum.yield.general.temp} and \eqref{eq.diff.limited}
results in:

\begin{equation}\label{eq:bushueva}
  f(T) = \left( p_1 + p_2\ T/\eta(T) \right)^{-1}
\end{equation}
where $p_1$ and $p_2$ are temperature-independent parameters.

The assumption that the temperature effect on $q$ is related to a
diffusion limited reaction may be accepted for solvent exposed
chromophores where collisional quenching by solvent and solutes is
possible. For buried aromatic residues on the other hand, one
would expect more complex mechanisms. Although it has been shown
in some cases \citep{gavish1979viscosity,beece1980solvent} that
protein dynamics is coupled to solvent viscosity, the issue is complicated both for folded \citep{ansari1992role} and unfolded \citep{soranno2012quantifying} proteins.
Thus, Eq.~\eqref{eq:bushueva} should perhaps be thought of as an
empirical relationship that may in some cases describe the
temperature dependence of fluorescence intensities \citep{lindorff2001surprisingly}. Indeed, another possibility is to use a simple, empiric quadratic equation to capture non-linear effects
\citep{saini2010relationship,hamborg2020}.

As for denaturant induced protein unfolding, we combine the above to arrive at a final expression that can be used for fitting heat induced protein unfolding measured by a spectrometic method. For the native state (here called $N$) where the aromatic residues are buried in the hydrophobic core of the protein it is often sufficient to use a linear temperature dependence:

\begin{equation}\label{eq:N.baseline.te}
    \overline{\alpha}_N\left(T\right)=\overline{\alpha}_{N,T_{ref}}+{\gamma}_{N}\left(T-T_{ref}\right)
\end{equation}
For the unfolded state (here called $U$) a linear temperature dependence is typically sufficient if the signal is measured by CD spectroscopy. For a signal measured by fluorescence spectroscopy a quadratic temperature dependence may be used as discussed above:

\begin{equation}\label{eq:U.baseline.te}
    \overline{\alpha}_U\left(T\right)=\overline{\alpha}_{U,T_{ref}}+{\gamma}_{1,U}\left(T-T_{ref}\right)+{\gamma}_{2,U}\left(T-T_{ref}\right)^2
\end{equation}
Inserting Eqs.~\eqref{eq:N.baseline.te} and \eqref{eq:U.baseline.te} in Eq.~\eqref{eq:santoro} and using $K=\exp(-\Delta_rG_U(T)/RT)$ we get:

\begin{equation}\label{eq.heat.unf}
    \overline{\alpha}\left(T\right)= \frac{\left(\overline{\alpha}_{N,T_{ref}}+{\gamma}_{N}\Delta T\right) +\left(\overline{\alpha}_{U,T_{ref}}+{\gamma}_{1,U}\Delta T+{\gamma}_{2,U}\Delta T^2\right) \exp{\left(\frac{-\Delta_rG_U(T)}{RT}\right)}} {1+\exp{\left(\frac{-\Delta_rG_U(T)}{RT}\right)}}
\end{equation}
where $\Delta T = T - T_{ref}$ and $\Delta_rG_U(T)$ given by Eq.~\eqref{eq:constant.deltaCp.2}.

\begin{figure}[tbp]
  \centering
  \includegraphics[height=5cm]{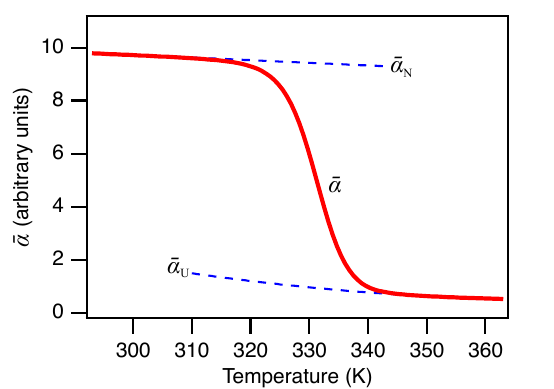}
  \caption{
            Heat induced unfolding. The red curve is synthetic experimental data. The dashed blue curves are estimates of $\overline{\alpha}_N$ and $\overline{\alpha}_U$ extrapolated from regions with almost pure $N$ and $U$ into the transition zone. Note that $\overline{\alpha}_U$ is curved. The curves were simulated using Eqs.~\eqref{eq:constant.deltaCp.2}, \eqref{eq:N.baseline.te}, \eqref{eq:U.baseline.te} and \eqref{eq.heat.unf} with $\overline{\alpha}_{N,T_{ref}}$ = 10, $\gamma_N$ = -0.01 K^{-1}, $\overline{\alpha}_{U,T_{ref}}$ = 3, $\gamma_{1,U}$ = -0.05 K^{-1}, $\gamma_{2,U}$ = 0.00025 K^{-2}, $\Delta_rH(T_m)$ = 340 kJ~mol^{-1}, $\Delta_rC_p$ = 10 kJ mol^{-1}K^{-1}, $T_m$ = 331 K and $T_{ref}$ = 273 K.}
  \label{fig:extrapolate-temp}
\end{figure}

Eq.~\eqref{eq.heat.unf} has been used in the literature to analyse temperature unfolding followed in particular by CD spectroscopy \citep{greenfield2013}. In addition to the four or five parameters describing the baselines, three thermodynamic parameters, $\Delta H_U(T_m)$, $\Delta _rC_{p,U}$ and $T_m$ are obtained from fits to the equation. The values obtained for the thermodynamic parameters depend on the change in the signal in a narrow interval around $T_m$. Often both $T_m$ and $\Delta H_U(T_m)$ can be determined from the analysis with a reasonable accuracy. However, $\Delta _rC_{p,U}$ can generally not be reliably fit from spectroscopic measurements of thermal unfolding \citep{zweifel2001studies}. This parameter divided by $T$ gives the curvature in how $\Delta _rG_U$ changes with $T$ (Eq.~\eqref{eq:deltaG.curvature}), which in the interval around $T_m$ is small compared to the slope of $\Delta _rG_U$. It is thus generally advised not to fit $\Delta _rC_{p,U} = 0$ from such data unless the data is of very high quality and reveals clear curvature of $\Delta _rG_U$. Instead, it is advisable to set $\Delta _rC_{p,U} = 0$ in Eq.~\eqref{eq:constant.deltaCp.2}, or to use a predicted value \citep{geierhaas2007bppred}, when fitting to Eq.~\eqref{eq.heat.unf}.  $\Delta _rC_{p,U}$ is better determined from a DSC experiment or from a series of CD or fluorescence detected heat induced unfolding experiments where $\Delta _rG_U$ is changed by adding denaturant (see below).

\subsection*{Differential scanning calorimetry}
Differential scanning calorimetry (DSC) is another method which
can be used to study protein stability \citep{freire1995differential,johnson2013}.
A detailed description of how to perform and analyse DSC is outside the
scope of this review, but we provide a basic description so that it can be compared 
to measurements e.g. of thermal unfolding monitored by spectroscopic measurements.

The measured quantity in a DSC experiment is the heat capacity of the
system, which thus does not rely on the presence of fluorophores or other
probes to measure stability. For this and other reasons DSC
complements well the techniques described so far. In a typical DSC
setup, two cells, one containing the protein sample and the other a
reference, are heated by some specific heating rate (the scan
rate). Through an electric feedback loop the temperature
difference between the two cells is kept very close to zero. The power
used to keep the temperature identical in the two cells is
directly related to the difference in heat capacity between the
two cells. Since the difference between the two cells is the
presence of protein in the sample, one directly measures the heat
capacity of the protein.

\begin{figure}[tbp]
  \centering
  \includegraphics[height=5cm]{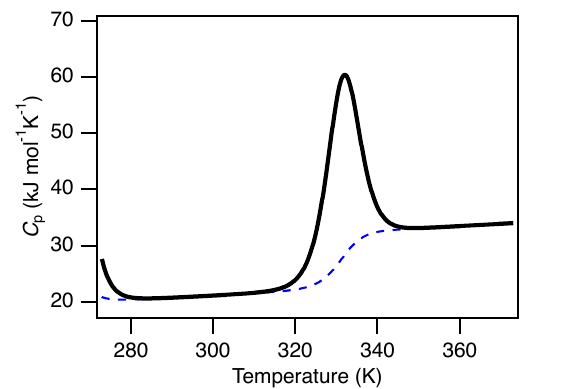}
  \caption{
            Synthetic data representing a calorimetric scan showing one peak centred at 331 K (58 $^\circ$C) corresponding to the heat unfolding of a protein. At temperatures below 280 K the beginning of a cold denaturation transition is just appearing. The black curve represents the synthetic experimental data. The blue curve is the baseline. The curve was simulated using Eq.~\eqref{eq.DSC.meas} with the parameters $\Delta_rH(T_m)$ = 340 kJ mol^{-1}, $\Delta_rC_{p,U}$ = 10 kJ mol^{-1}K^{-1}, $T_m$ = 331 K and with a slope of both the pre- and post-transition baselines of 0.04 kJ mol^{-1}K^{-2}.}
  \label{fig:dsc}
\end{figure}

A hypothetical DSC curve is shown in Fig.~\ref{fig:dsc}. Like the
other protein unfolding experiments discussed so far, it consists
of three regions. First (above 280K) a pre-transition baseline, then a
transition region and finally a post-transition region. The major
feature in the curve is a large peak associated with the unfolding
of the protein. Before this peak, the baseline represents the heat
capacity of the native protein. After the peak the baseline
represents the heat capacity of the denatured protein. Finally, we note that at the lowest temperatures in the DSC data there is evidence of cold-denaturation in line with the temperature dependency of $\Delta_r G$ with these thermodynamic parameters (Fig.~\ref{fig:temperature-profile}).

As discussed in the section on heat induced unfolding, $\Delta_rC_{p,U}$ for proteins are positive and therefore the post-transition baseline will lie
higher than the pre-transition baseline. It is in general a good approximation to assume that
$\Delta_rC_{p,U}$ is temperature independent, which means
that the pre- and post-transition baselines are parallel, although more complex baselines may be used on high quality data \citep{privalov2007baseline}.

DSC experiments are generally
performed by first making a temperature scan of the background without protein and then subtracting this from a second temperature scan of the protein sample. The measured heat consumption is composed of population-weighted temperature-dependent heat capacities of the folded, $C_{p,N}(T)$ and unfolded, $C_{p,U}(T)$ states (the baseline) and an extra heat contribution, $C_{p,unf}$ resulting from the unfolding of the protein. Using the assumption that the pre- and post-transition baselines are parallel with a common slope, $\gamma$, we get: 
\begin{align}
    &C_{p,N}(T)=C_{p,N}(T_m)+\gamma(T-T_m)\label{eq:DSC.Nbas}\\
    &C_{p,U}(T)=C_{p,U}(T_m)+\gamma(T-T_m)\label{eq:DSC.Ubas}
\end{align}
As the temperature increases the fraction of protein in the native state, $p_N$, changes. The heat needed to drive this change is $\Delta_rH_U$ and the heat capacity associated with unfolding is thus:
\begin{equation}\label{eq:DSC.excess}
    C_{p,unf}(T)=\Delta_rH_U(T)\frac{\partial{p_N}}{\partial{T}}
\end{equation}
For a two-state reaction, the derivative in Eq.~\eqref{eq:DSC.excess} can be evaluated by using $p_N=1/(1+K(T))$ which gives:
\begin{equation}\label{eq:DSC.dp.dT}
    \frac{\partial{p_N}}{\partial{T}}=\frac{\partial{(1/(1+K(T)))}}{\partial{T}}=\frac{\exp{(-\Delta_rG_U(T)/RT)}}{(1+\exp{(-\Delta_rG_U(T)/RT)})^2}\frac{\Delta_rH_U(T)}{RT^2}
\end{equation}
where $\Delta_rH_U(T)=\Delta_rH_U(T)+\Delta_rC_{p,U}(T)(T-T_m)$ and $\Delta_rG_U(T)$ is given by Eq. \eqref{eq:constant.deltaCp.2}. Combining Eqs. \eqref{eq:DSC.Nbas}-\eqref{eq:DSC.dp.dT} gives a final expression for the heat capacity measured in a DSC experiment of a protein following two-state folding:
\begin{equation}\label{eq.DSC.meas}
    \begin{aligned}
    C_{p,meas}(T) & = p_NC_{p,N}(T)+p_UC_{p,U}(T)+\Delta_rH_U(T)\frac{\partial{p_N}}{\partial{T}}\\
    & = C_{p,N}(T)+(1-p_N)\Delta_rC_{p,U}+\frac{\exp{(-\Delta_rG_U(T)/RT)}}{(1+\exp{(-\Delta_rG_U(T)/RT)})^2}\frac{(\Delta_rH_U(T))^2}{RT^2}
    \end{aligned}
\end{equation}
In summary, the procedure for measuring protein unfolding by DSC is:
\begin{enumerate}
  \item  A blank scan is performed in order to correct for 
            small differences between the two cells.
  \item  A temperature scan of the sample \emph{vs}. reference is  performed.
  \item  The sample is rescanned in order to evaluate whether 
            the unfolding is reversible.

  \item  The blank scan is subtracted from the sample scan
  \item  One decides on appropriate functions for the pre- and post-transition baselines and how to join these (Above we used parallel straight lines and a two-state transition).
  \item  The measured data are fitted to Eq.~\eqref{eq.DSC.meas}.
\end{enumerate}

In Fig.~\ref{fig:dsc} we have used a simple two-state model to connect  the two parallel baselines according to the amount of
native and denatured protein present at that temperature \citep{takahashi1981thermal}. As mentioned above, the unfolding behaviour may be more complex and require other models for the baselines and non-two-state models to describe the unfolding reaction \citep{seelig2016dsc}. Because the choice of baselines are model dependent, they may affect the interpretation of the data. For example, it has been shown that different choices of baselines may affect
whether a system is thought to behave like a two-state system or not \citep{zhou1999calorimetric}.

\section*{Combining denaturant and heat induced unfolding}
So far we have derived expressions for how the observable $\overline{\alpha}$ changes with temperature at a constant denaturant concentration (Eq.~\eqref{eq:alpha.vs.D}) and how it changes with denaturant concentration at a constant temperature (Eq.~\eqref{eq.heat.unf}).  If heat induced unfolding of a protein is measured at a series of different denaturant concentrations, we need to extend the formalism to allow for a combined global analysis of all the data. The procedure is the same as we used above with the addition that we need to find expressions for how the parameters $K$, $\overline{\alpha}_N$ and $\overline{\alpha}_U$ vary when both $T$ and $D$ are changed \citep{zweifel2002,hamborg2020}.

\subsection*{$K$ as a function of denaturant and temperature}
The task here is to find an expression for $K_U\left(T,D\right)$ or $\Delta_rG_U\left(T,D\right)$. As $\Delta G$ is a state function, we can first evaluate the effect of changing the temperature and then at the new temperature evaluate the effect of changing the denaturant concentration: 
\begin{equation}
    \Delta_rG_U\left(T,D\right)=\int_{T_m}^{T}\left(\frac{\partial{\Delta_rG}_U}{\partial T}\right)_{D=0} \mathrm{d}D+\int_{0}^{D}\left(\frac{\partial{\Delta_rG}_U}{\partial D}\right)_T \mathrm{d}T
\end{equation}
As our reference point we have chosen $T = T_m$  and $D = 0$. From our previous discussion we have the solutions to these two integrals from Eq. \eqref{eq:constant.deltaCp.2} and Eq. \eqref{eq:LEM}, respectively, which give:

\begin{equation}\label{eq.dG.T.D}
    \Delta_rG_U\left(T,D\right)=\Delta_rH_U\left(T_m\right)\left(1-T/T_m\right)+\Delta_rC_{p,U}\left(T-T_m-T\ln\left(T/T_m\right)\right)-m\left(T\right)D
\end{equation}
where $T_m$ and $\Delta_rC_{p,U}$ are the values at $D = 0$. As discussed above we assume that $\Delta_rC_{p,U}$ is independent of temperature. The first integral is calculated at $D = 0$ and therefore it is not necessary to know how $\Delta_rC_{p,U}$ or $T_m$ varies with $D$. The second integral is evaluated at the temperature $T$ and we need to know the temperature dependence of $m$. In the LEM, $m$ is itself a change in free energy and is expected to have a temperature dependence similar to Eq. \eqref{eq:constant.deltaCp.2} \citep{chen1989,neira2004},

\begin{equation}
    m\left(T\right)=m_{\Delta H_i}\left(T_{ref}\right)\left(1-T/T_{ref}\right)+m_{\Delta C_{p,i}}\left(T-T_{ref}-T\ln\left(T/T_{ref}\right)\right)
\end{equation}
where $m_{\Delta H_i}$ and $m_{\Delta C_{p,i}}$ are the enthalpy and heat capacity associated with the preferential interaction of denaturant with the unfolded state at the reference temperature $T_{ref}$. Several studies have determined $\Delta_rG_U\left(D\right)$ at multiple temperatures from denaturant induced unfolding experiments and consequently been able to evaluate the temperature dependence of $m$. As far as we are aware, curvature in $m(T)$ has only been observed in a single case  \citep{zweifel2002} suggesting that $m_{\Delta C_{p,i}}$ in general is very small.  The variation of $m$ with $T$ is overall small, and in some cases $m$ was approximated to be independent of temperature \citep{chen1989,agashe1995,hollien1999,hamborg2020}. In other cases, however, $m$ decreases linearly with temperature \citep{makhatadze1992,nicholson1996,dekoster1997,felitsky2003,neira2004,amsdr2019urea} and can be approximated by:

\begin{equation}
    m\left(T\right)=m_{\Delta H_i}\left(T_{ref}\right)\left(1-T/T_{ref}\right)
\end{equation}
Depending on how $m$ varies with $T$, four to six parameters are thus needed to describe $\Delta_rG_U\left(T,D\right)$.

\subsection*{$\overline{\alpha}_N$ and $\overline{\alpha}_U$ as functions of denaturant and temperature}
\begin{figure}[tbp]
  \centering
  \includegraphics[height=5cm]{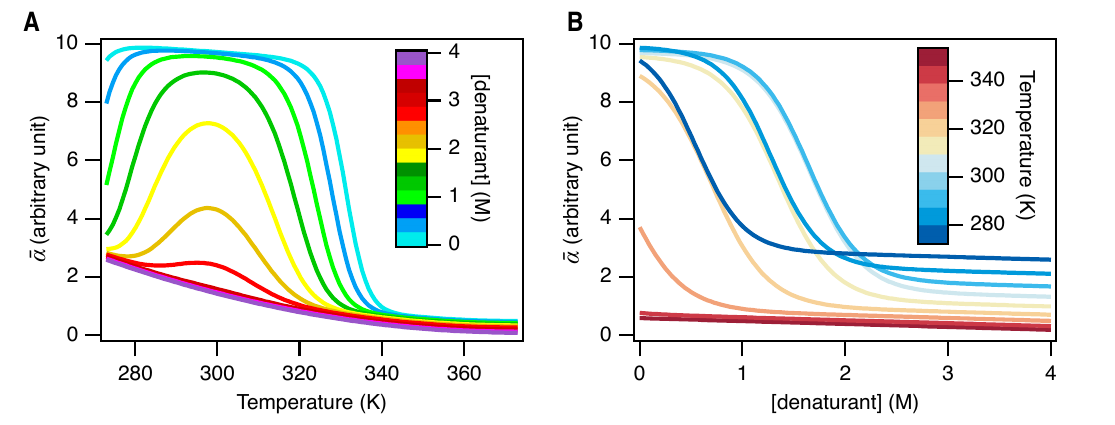}
  \caption{
            Combined temperature and denaturant induced unfolding. (A) Synthetic temperature unfolding curves at denaturant concentration from 0 to 4 M as indicated by the colour scale. (B) Synthetic denaturant unfolding curves at temperatures from 273 K to 353 K as indicated by the colour scale. The curves were simulated using Eqs.~\eqref{eq.dG.T.D} and \eqref{eq.heat.denat.unf} with $\overline{\alpha}_{N,0,T_{ref}}$ = 10, $\beta_N$ = -0.05 M^{-1}, $\gamma_N$ = -0.01 K^{-1}, $\overline{\alpha}_{U,0,T_{ref}}$ = 3, $\beta_U$ = -0.01 M^{-1}, $\gamma_{1,U}$ = -0.05 K^{-1}, $\gamma_{2,U}$ = 0.00025 K^{-2}, $\Delta_rH(T_m)$ = 340 kJ mol^{-1}, $\Delta_rC_p$ = 10 kJ mol^{-1}K^{-1}, $T_m$ = 331 K and $T_{ref}$ = 273 K.}
  \label{fig:extrapolate-2D}
\end{figure}

Taking the outset from Eqs.~\eqref{eq:N.baseline.D}, \eqref{eq:U.baseline.D}, \eqref{eq:N.baseline.te} and \eqref{eq:U.baseline.te} we know how the pre- and post-transition baselines change with denaturant at constant temperature, and how they change with temperature at constant denaturant. The question now is if the slopes of the baselines in the denaturant dimension are temperature dependent or if the slopes of the baselines in the temperature dimension are denaturant dependent. If we assume a linear change of the slopes, we can express the plane $\overline{\alpha}_i(T,D)$ for state $i$ as:
\begin{equation}\label{eq.baseplane}
    \begin{aligned}
        \overline{\alpha}_i\left(T,D\right) &=\overline{\alpha}_i\left(T_{ref},0\right)+\left({\beta_{i}}+{\beta'_{i}}{\Delta}T\right)D+\left({\gamma_{i}}+{\gamma'_{i}}D\right){\Delta}T\\
        &=\overline{\alpha}_i\left(T_{ref},0\right)+{\beta_{i}}D+{\gamma_{i}}{\Delta}T+\left({\beta'_{i}}+{\gamma'_{i}}\right)D{\Delta}T
    \end{aligned}
\end{equation}
where $\Delta T=\left(T-T_{ref}\right)$. Only few studies in the literature have fitted the pre- and post-transition planes, and to our knowledge in none of these the last $D{\Delta}T$-dependent term was included to obtain satisfactory fits of the experimental data \citep{nicholson1996,hollien1999,neira2004,hamborg2020}. We thus proceed by assuming that the $D{\Delta}T$ crossterm is negligible, that the pre-transition plane increases linearly with both $T$ and $D$ (Eqs.~\eqref{eq:N.baseline.D} and \eqref{eq:N.baseline.te}) and that the post-transition plane increases linearly with $D$ (Eq.~\eqref{eq:U.baseline.D}) and quadratically with $T$ (Eq.~\eqref{eq:U.baseline.te}). Combining all this with Eqs.~\eqref{eq:santoro}, \eqref{eq:LEM} and \eqref{eq:constant.deltaCp.2} we get an expression to explain and fit a spectrometric signal for protein unfolding in the two-dimensional $(T,D)$ space:

\begin{equation}\label{eq.heat.denat.unf}
    \overline{\alpha}\left(T,D\right)= \frac{\left(\overline{\alpha}_{N,0,T_{ref}}+{\beta_{N}}D+{\gamma_{N}}\Delta T\right) +\left(\overline{\alpha}_{U,0,T_{ref}}+{\beta_{U}}D+{\gamma_{1,U}}\Delta T+{\gamma_{2,U}}\Delta T^2\right) \exp{\left(\frac{-\Delta_rG_U(T,D)}{RT}\right)}} {1+\exp{\left(\frac{-\Delta_rG_U(T,D)}{RT}\right)}}
\end{equation}
where $\overline{\alpha}_{i,0,T_{ref}}$ is the spectroscopic signal at $T_{ref}$ in the absence of denaturant, $\Delta T = T - T_{ref}$ and $\Delta_rG_U(T,D)$ given by Eq.~\eqref{eq.dG.T.D}. Such two-dimensional unfolding experiments have been used to study the stability of very stable designed proteins \citep{jacobs2016design} and for the analyses of variants with widely different stabilities \citep{zutz2020dual}. Examples of series of unfolding curves in both the temperature and denaturant dimension are shown in Fig.~\ref{fig:extrapolate-2D}.

\section*{Equilibrium two-state folding}
The discussions above generally assume a two-state behaviour, and we have shown how 
any transition may \emph{formally} be treated as such, at least in
equilibrium experiments. It is, however, also clear that if the system
does not behave like a two-state system, but is treated as such,
the baselines would have highly non-trivial shapes. This is clear from Eq.~\eqref{eq:averagealphas}
which show that unless the subpopulations within each state behave smoothly as a function
of a perturbing variable (e.g. temperature) then $\overline{\alpha}_N$ and $\overline{\alpha}_U$ could
be highly non-linear. Thus, it becomes important to consider how much the system behaves more like a
thermodynamic system with two separate states, and where the properties 
of these states are only mildly dependent on the conditions so that they can be considered mostly fixed. This in turn highlights the baselines as potential dumping ground for errors that
arise when fitting a multi-state transition to a two-state model, and hence again 
highlights the importance of sensible choices of baselines and examining how well
they fit the data.

Focussing on equilibrium properties, a number of tests have been put forward  to address
the issue of `two-state or not' \citep{lumry1966validity}. These
tests include the analysis of the temperature dependence of
thermodynamic parameters, estimation of parameters using several
independent techniques and perhaps the most well-known test of
all, the `calorimetric criterion'. This criterion states that for
a two-state reaction, the unfolding reaction-enthalpy which is
obtained by integrating the area between a DSC peak and the merged
baselines, should be identical to the enthalpy obtained from the
van't Hoff equation (Eq.~\eqref{eq:vant.hoff}) and the temperature
dependence of the equilibrium constant determined in the same
experiment. It has been shown \citep{zhou1999calorimetric} that for
\emph{any} DSC experiment, the calorimetric criterion can be
fulfilled by some `clever' choice of baseline.
Indeed, this is not just a philosophical issue, as evidenced by the substantially
different interpretations of equilibrium unfolding experiments depending in
part of the choices of baselines and models \citep{sadqi2006atom,ferguson2007analysis,zhou2007analysis,sadqi2007analysis}.

\section*{Conclusions}
We have here reviewed theory, equations and approaches to study protein stability by experiments.
Rather than being comprehensive, we have focused on linking the fundamental properties of molecules with
experimental observables. We have focused on equilibrium folding experiments of reversible folding, and discussed general procedures for
thinking about these in terms of states, average properties within states, and perturbing the populations of these states 
by e.g. denaturants or temperature. This procedure involves two key steps. First, one needs to model how
the relative population of the states change as a function of the perturbing variable(s), and we have discussed 
this extensively for denaturants and temperature.
Second, because the states are themselves ensembles, and because the spectroscopic 
probes are sensitive to the conditions, there can be changes to the signals even within the states
making it important also to model the baselines.
In most current analyses of experiments, standard choices are made for these steps. E.g.
nearly all denaturant-induced unfolding experiments are modelled using the LEM for the stability
and linear baselines for fluorescence or CD spectroscopy. We hope, however, the reader will 
have learnt more about the origin, assumptions and potential pitfalls that are involved in such approaches.

% =======================================================
%           Acknowledgements
% =======================================================
\section*{Acknowledgments}
We acknowledge support for the Protein Interactions and Stability in Medicine and Genomics centre (PRISM; NNF18OC0033950) and Protein OPtimization (POP; NNF15OC0016360), both funded by the Novo Nordisk Foundation.
% =======================================================
%           References
% =======================================================
\bibliography{references}

\end{document}